\documentclass{aastex}

\begin{document}
\title{Dwarf Novae with Newly Determined Parallaxes:
Model Analyses of VY Aquari, RU Pegasi, and T Leonis}

\author{R.T. Hamilton and E.M. Sion }
\affil{Department of Astronomy and Astrophysics,
Villanova University,
Villanova, PA 19085,
e-mail: ryan.hamilton@villanova.edu,
       edward.sion@villanova.edu}

\begin{abstract}

Using newly determined parallaxes for dwarf novae, we have derived
outburst accretion rates for VY Aqr, RU Peg and T Leo and for T Leo during
quiescence.  The two short-period dwarf novae, VY Aqr and T Leo, show good
agreement with optically thick steady-state accretion disks in outburst,
whereas RU Peg shows a significant departure from a steady-state disk.  
We have determined that the white dwarf in T Leo has T$_{eff} = 16,000 \pm
1000$ K, a value consistent with long term compressional heating when
gravitational wave emission drives mass transfer. The white dwarf in T Leo
has a temperature in the same narrow range as other WZ Sge-like dwarf
novae.
 
\end{abstract}


\section{Introduction}
       
Distances from accurate parallaxes for cataclysmic variables are
critically important to model-fitting determinations of the mass accretion
rates, accreting white dwarf temperatures, and the identification of the
source of far UV (FUV) continuum during quiescence and outburst states.
Recently, Thorstensen (2003) and Harrison et al. (2004) have presented new
parallaxes for dwarf novae from ground-based and {\it Hubble} Fine
Guidance Sensor (FGS) determinations respectively. These parallaxes offer
the opportunity to more accurately analyze the {\it IUE} archival spectra
of dwarf novae during outburst and quiescence. By knowing the distance,
model-derived accretion rates and white dwarf properties are known more
reliably. Only with good distances can it be known what fraction of the
FUV flux is being contributed by what radiating component, be it the white
dwarf, the boundary layer or the accretion disk. We have selected three
dwarf novae systems with recently measured parallaxes: VY Aquari, an SU
UMa system below the period gap; RU Peg, a U Gem system above the period
gap; and T Leo, an SU UMa system below the period gap. A summary of these
systems and their associated properties can be found in Table 1.

\section{The IUE Archival Observations}

All three systems were observed with {\it IUE} and have usable spectra
taken with the Short Wavelength Prime (SWP) camera covering the wavelength
region from 1170 - 2000 \AA.  We obtained the spectra through the MAST
{\it IUE} archive, and then applied the Massa \& Fitzpatrick (2000) flux
calibrations to correct the NEWSIPS data.  For VY Aquari, a characteristic
outburst spectrum, whose time placement is shown in the AFOEV light curve
in Figure 1, was chosen for analysis.  The quiescent state of VY Aqr has
already been analyzed by Sion et al. (2003) using {\it HST} STIS spectra.  
For the analysis of RU Peg, we chose two usable outburst spectra in the
MAST {\it IUE} archive: one near the peak of outburst (SWP15062), and one
occurring on the decline from an outburst (SWP15079) as seen in the AAVSO
lightcurve in Figure 2.  Quiescent {\it IUE} spectra of RU Peg has already
been analyzed by Sion \& Urban (2002).  The {\it IUE} spectra of T Leo
covers both the outburst and quiescent states, with outburst spectra
having been previously analyzed by Belle et al. (1998) and the quiescent
spectra available having no previous model analysis.  We present a new
analysis of T Leo's outburst spectrum utilizing an updated grid of
steady-state accretion disk models and white dwarf photospheres.  The time
placement of the outburst and quiescent spectra chosen for our analysis of
T Leo are shown in the AAVSO lightcurve in Figure 3.  The outburst
spectrum is seen to occur just after peak outburst, and the quiescent
spectrum shown just after the return to optical quiescence.

A summary of the {\it IUE} spectra can be found in Table 2.  We have
listed the system name, SWP image number of the specific observation,
date, and exposure length.  All were obtained through the large aperture
of {\it IUE} with the exception of SWP21720 as indicated in the table.

\section{Synthetic Spectral Fitting Technique}

We pursued three avenues in modeling the observations with synthetic
spectra: an accretion disk model alone, a single white dwarf photosphere,
or a combination of the two.

For the accretion disks, we adopted models from Wade \& Hubeny (1998).  
Using {\it IUEFIT}, a $\chi^{2}$ minimization routine, the model disk was
scaled and fit to the spectrum.  The fitting scale factor can then be
shown to be related to the white dwarf distance in pc though
$d=100/\sqrt{(S)}$ where {\it S} is the scale factor given by the {\it
IUEFIT} routine, {\it d} is the system distance in pc, and the factor of
100 arises from the fact that the theoretical disk fluxes are normalized
to a distance of 100 pc.  Using this approach we have two parameters in
determining the goodness of a fit: a minimum $\chi^{2}$ and a scale-factor
distance to compare with the parallax distance.

For single photospheres, we used the codes {\it TLUSTY} (Hubeny 1988) in
conjunction with {\it SYNSPEC} with {\it ROTIN3} (Hubeny \& Lanz 1995) to
generate synthetic photosphere spectra convolved with the {\it IUE}
instrumental profile.  We generated a new grid of solar abundance models,
covering a range of temperatures from 15,000 - 50,000 K in increments of
1000 K with $\log(g)$ ranging from 7.0 to 8.6 in increments of 0.2.  This
grid of models was then applied to the observations using {\it IUEFIT}.
The scale factor {\it S} for the photosphere fits is related to the radius
of the white dwarf given by $R_{wd}=(\frac{d}{1000})\sqrt{(S)}R_{\sun}$,
where $R_{\sun}$ is the radius of the Sun ($6.96\times 10^{10}$ cm), {\it
d} is the known system distance in pc, and the factor of 1000 arises from
the fact that the theoretical photosphere fluxes are normalized to a
distance of 1000 pc.  This again gives two parameters for determining the
goodness of the fit, $\chi^{2}$ and the white dwarf radius computed using
the parallax distance.

The best fitting accretion disk and photosphere models were also combined
in an attempt to achieve a better fit.  We used our code {\it DISKFIT},
which allows us to vary the accretion rate linearly between 0.1 and 10.  
This effectively alters the contribution of the disk to the observed flux,
giving the underlying white dwarf a greater or lesser flux contribution
depending on the ratio.  The fitting routine computes a scale factor that
is related to the system distance as given by $d=100/\sqrt{(S)}$ with
distance given in pc.

\section{Model Fitting Results}

\subsection{VY Aquari in Outburst}

For VY Aquari, the best fitting optically thick steady-state accretion
disk model yields an accretion rate of $10^{-9}$ M$_{\sun}$ yr$^{-1}$.  
This compares to the prediction of a time averaged accretion rate of
approximately $5\times 10^{-10}$ M$_{\sun}$ yr$^{-1}$ for OY Car, an SU
UMa system with an orbital period very similar to that of VY Aqr, given by
Patterson (1984). 

The white dwarf mass and disk inclination used in this best fit (shown in
Figure 4) are $0.55$ M$_{\sun}$ and 41\degr\ respectively, giving a
scale-factor distance of 93 pc.  This is in very good agreement with the
parallax distance of $97 \pm 13$ pc obtained by Thorstensen (2003). Notice
that this best fit agrees well with the observations, and does not show
any significant deviations from a steady-state disk longward of 1600 \AA.

Note that in the case of VY Aqr, no previous mass determinations existed
and the inclination was uncertain. In our fitting therefore, we examined
the widest range of $M_{wd}$ and the inclination i. Since the
goodness-of-fit is sensitive to both $M_{wd}$ and i, it is most helpful to
know or constrain one of the two parameters. The fits are based upon both
the continuum slope and Lyman Alpha wings.  However, an important
additional clue was provided by the depth of the Lyman alpha profile
before it is reversed by the geocoronal emission. This allowed us to rule
out three models having different M$_{wd}$ giving an acceptable distance and
$\chi^{2}$ values close to the best-fitting one.

\subsection{RU Peg in Outburst}

For RU Peg, our best fitting steady-state optically thick disk model
yields an accretion rate of $10^{-9}$ M$_{\sun}$ yr$^{-1}$ with a white
dwarf mass of 1.21 M$_{\sun}$ and an inclination of 41\degr\ (shown in
Figure 5).  This accretion rate is within a factor of 10 lower than that
of Patterson (1984).  The resulting scale factor distance of 278 pc gives
excellent agreement with the parallax distance of $282 \pm 20$ pc by
Harrison et al. (2004).  However, there is a significant deviation from a
steady-state disk flux distribution in the observed spectrum longward of
1600 \AA\ that could not be accounted for by a single disk, white dwarf
photosphere, or combination of the two.  The poor agreement between the
accretion disk models and the outburst spectrum also occurred for the
fitting of the outburst spectrum (SWP15062) obtained two days closer to
peak outburst.

\subsection{T Leo in Outburst}

For T Leo in outburst, we found that the spectrum is best fit by a
steady-state optically thick disk with an accretion rate of $10^{-8}$
M$_{\sun}$ yr$^{-1}$, a white dwarf mass of 0.35 M$_{\sun}$, and an
inclination of 60\degr\ (shown in Figure 6).  The resulting scale factor
distance of 113 pc is close to the parallax distance of $101 \pm 12$ pc by
Thorstensen (2003). It is interesting that the steady-state disk fit to T
Leo, like the fit to VY Aqr, shows good agreement with the observations.
There is no significant deviation from a steady-state flux distribution as
is seen in RU Peg.

\subsection{ T Leo in Quiescence}

For T Leo in quiescence, we found that the spectrum is not best fit by
either a single disk or white dwarf photosphere but by a combination of
the two.  Using the mass and inclination obtained from the best fit to the
outburst spectrum, we found that the quiescent spectrum is best fit by the
combination of a white dwarf with T$_{eff}$ of 16,000 K and $\log(g)$ of
7.5, an accretion disk with a white dwarf mass of 0.35 M$_{\sun}$, an
inclination of 60\degr\, and a combined accretion rate of $6\times
10^{-11}$ M$_{\sun}$ yr$^{-1}$. We estimate the error in our temperature
determination to be 1000 K.  This accretion rate is within a factor of 10
of Patterson's (1984) time averaged rate of $10^{-11}$ M$_{\sun}$
yr$^{-1}$.  The best fit (shown in Figure 7) also gives a scale-factor
distance of 101 pc, again in excellent agreement with the parallax
distance of $101 \pm 12$ pc by Thorstensen (2003). Our finding that the
white dwarf dominates the flux in the FUV wavelength range during
quiescence is consistent with the fact in the short period, large outburst
amplitude long recurrence time dwarf novae such as WZ Sge or AL Com during
quiescence, the accretion disk is not expected to be the major contributor
to the flux (Szkody et al. 2002).

\section{Conclusions}

We have derived accretion rates for 3 dwarf novae in outburst, and have
provided the first temperature determination of the white dwarf in T Leo
derived from the analysis of its quiescent {\it IUE} spectrum. For the
best accretion disk fits to the outburst spectra of VY Aquari and T Leo, a
steady-state optically thick accretion disk represents the observed FUV
energy distribution very well with no significant deviations. For RU Peg
in outburst, however, there is a large deviation from the observation
longward of 1600 \AA.  This implies an additional radiating component
(secondary star or hot spot?) or that the temperature distribution, T(r),
in the disk differs from the steady-state T(r).

The relatively low signal-to-noise of the IUE spectra introduces
considerable uncertainty in the accretion rates we have derived. By
comparison with the comparable quality IUE spectra of EM Cygni and CZ Ori
which were analyzed by Winter and Sion (2003) with a formal error
analysis, we estimate uncertainties in the accretion rates Log \.{M} of
$\pm 0.3$ for the three outburst spectra in this paper while for T Leo's
fainter spectrum in quiescence, we estimate a $\pm 0.5$ uncertainty in Log
\.{M}.

The temperature of the white dwarf in T Leo is consistent with the
predicted range of temperature expected from long term compressional
heating at a rate of mass transfer driven by gravitational wave emission
as shown by Sion et al. (2003) and Townsley and Bildsten (2002).  As seen
in Table 3, the effective temperature of the T Leo white dwarf agrees with
the white dwarf effective temperatures in dwarf novae with similar orbital
periods. Recently, Vrielmann et al. (2004) provided evidence from T Leo's
XMM-Newton X-ray light curve that it might contain a magnetic white dwarf
and hence be the first superoutbursting intermediate polar (IP). They
found a 414s signal which could be due to the rotation of the white dwarf.
However, it is also possible this signal could be an ordinary QPO. If T
Leo is an IP, our temperature for the white dwarf is one of only three
white dwarf temperatures known in intermediate polars.

\acknowledgements

We thank the many observers at AAVSO and AFOEV. We acknowledge support of
this research by NSF Grant AST99-01955 and NASA ADP Grant NNG04GE78G, and
also support by the Delaware Space Grant College and Fellowship Program
(NASA Grant NGT5-40024).

%% Figures %%

\clearpage

\begin{figure} 
\plotone{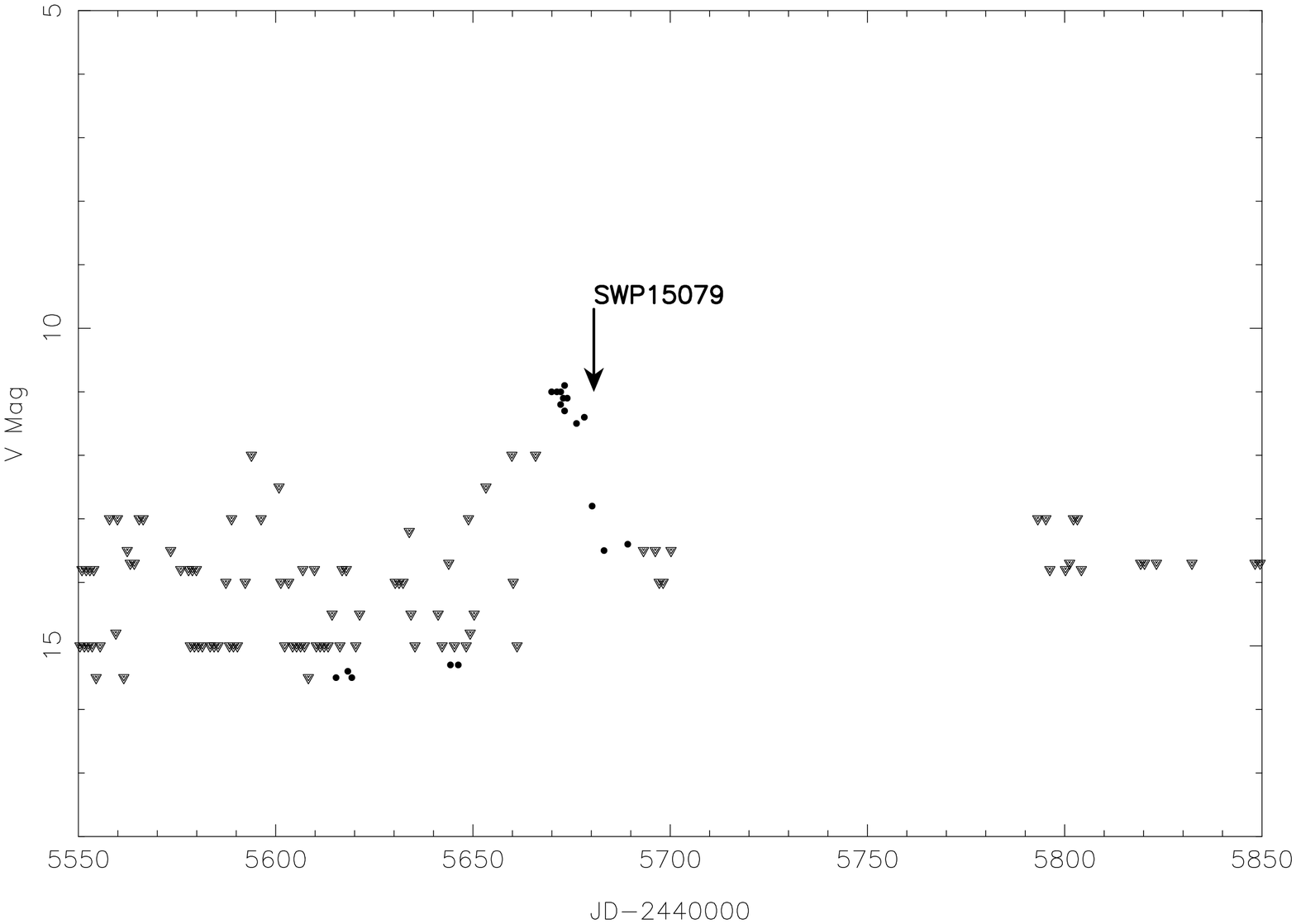} 
\caption{The AFOEV lightcurve of VY Aqr around the time
of the {\it IUE} observation shown, SWP21720.  Points marked by an inverted triangle are
upper limit estimates, whereas the solid points are photometric measurements.  Note the
large outburst amplitude ($\sim 5$ mag. ) characteristic of WZ Sge-like
systems.\label{fig01}} 
\end{figure}

\clearpage

\begin{figure} 
\plotone{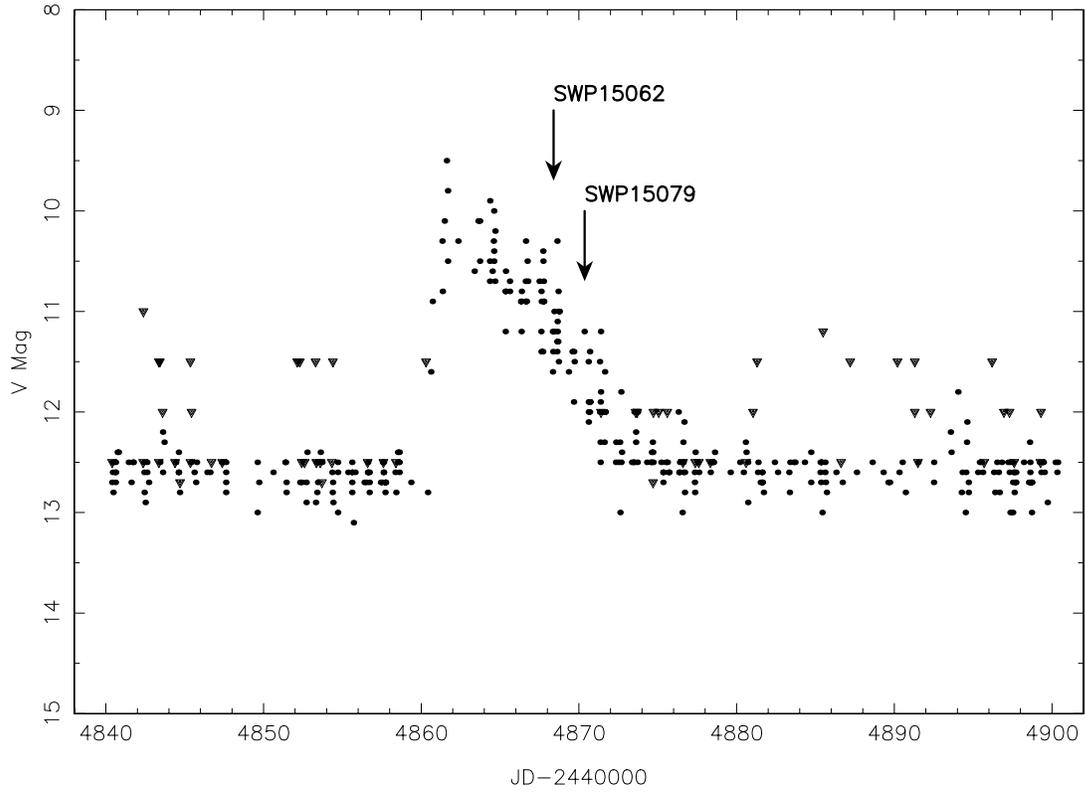} 
\caption{The AAVSO lightcurve of RU Peg around the time
of the {\it IUE} observation shown, SWP15079.  Points marked by an inverted triangle are
upper limit estimates, whereas the solid points are photometric measurements.\label{fig02}} 
\end{figure}

\clearpage

\begin{figure} 
\plotone{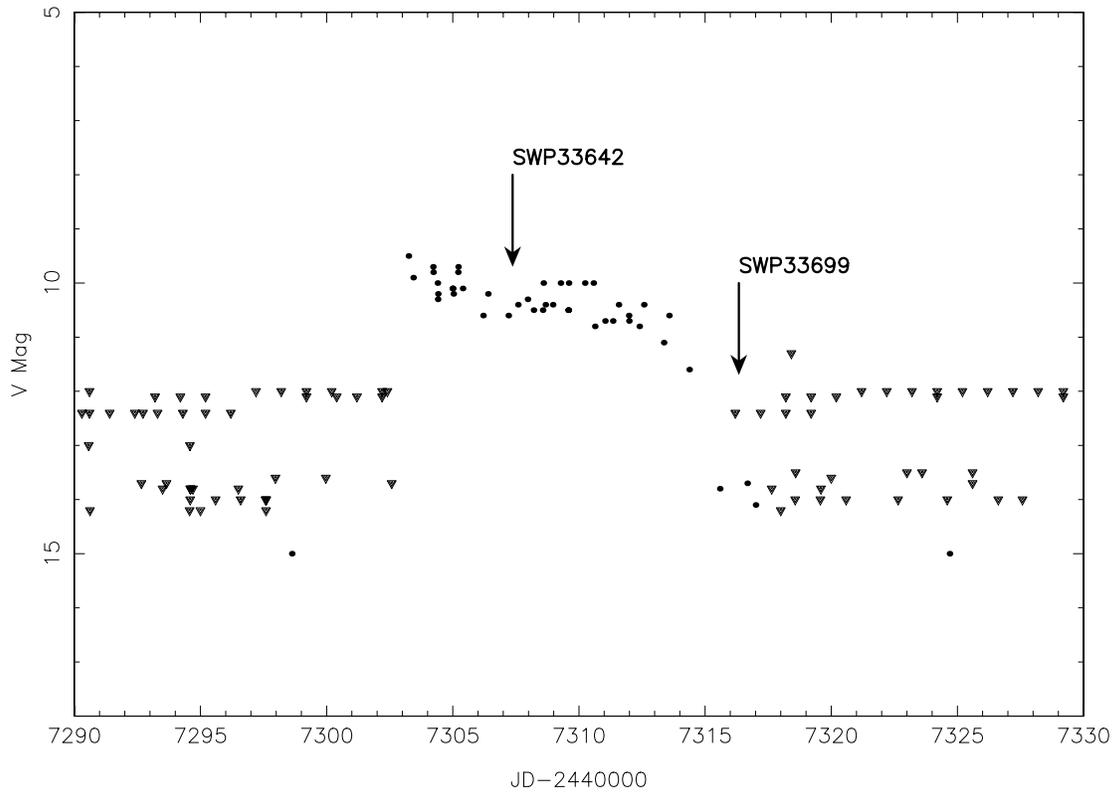} 
\caption{The AAVSO lightcurve of T Leo around the time
of the {\it IUE} observations shown, SWP33642 during outburst and SWP33699 during
quiescence.  Points marked by an inverted triangle are upper limit estimates, whereas the
solid points are photometric measurements.  Again note the large outburst amplitude
($\sim5$ mag) typical of WZ Sge-like systems.\label{fig03}} 
\end{figure}

\clearpage

\begin{figure}
\plotone{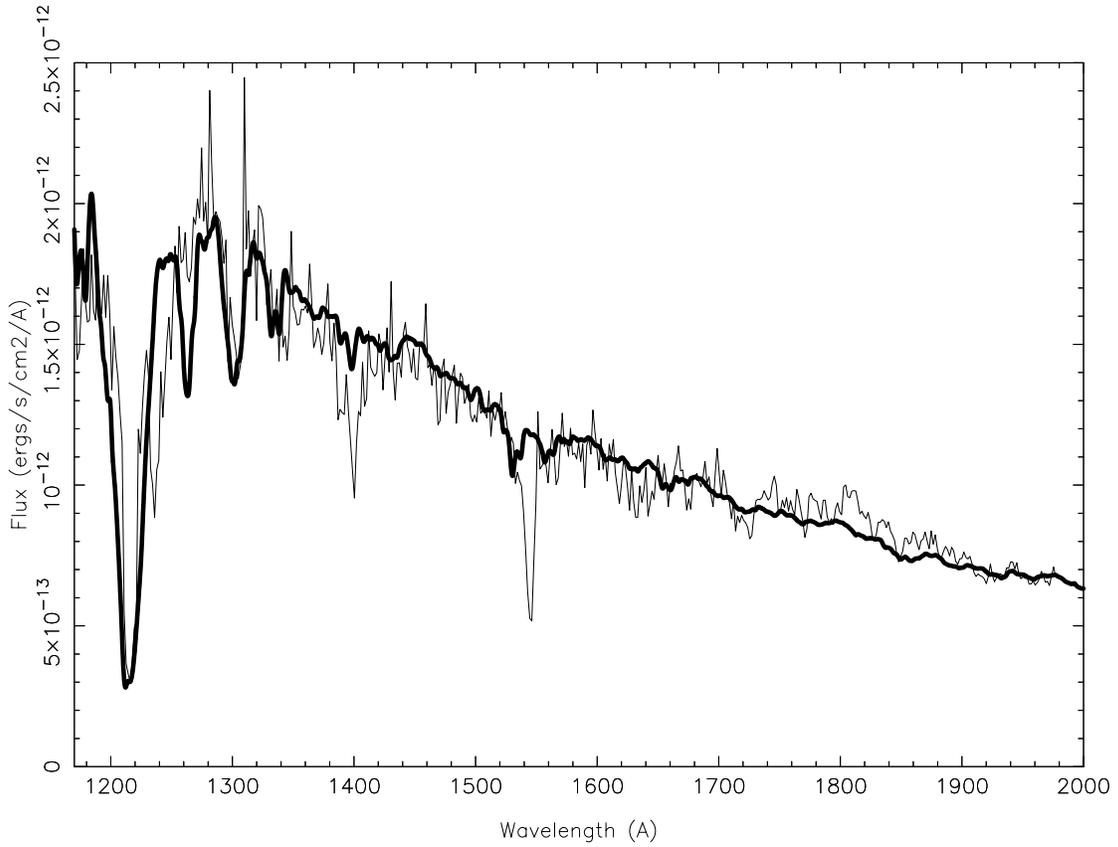}
\caption{The best fitting accretion disk to the {\it IUE} spectrum SWP21720 
for VY Aqr, having a white dwarf mass of 0.55 M$_{\sun}$, an accretion rate 
of $10^{-9}$ M$_{\sun}$ yr$^{-1}$, and an inclination of 41\degr.  This fit gives a 
scale-factor distance of approximately 93 pc.\label{fig04}}
\end{figure}

\clearpage

\begin{figure}
\plotone{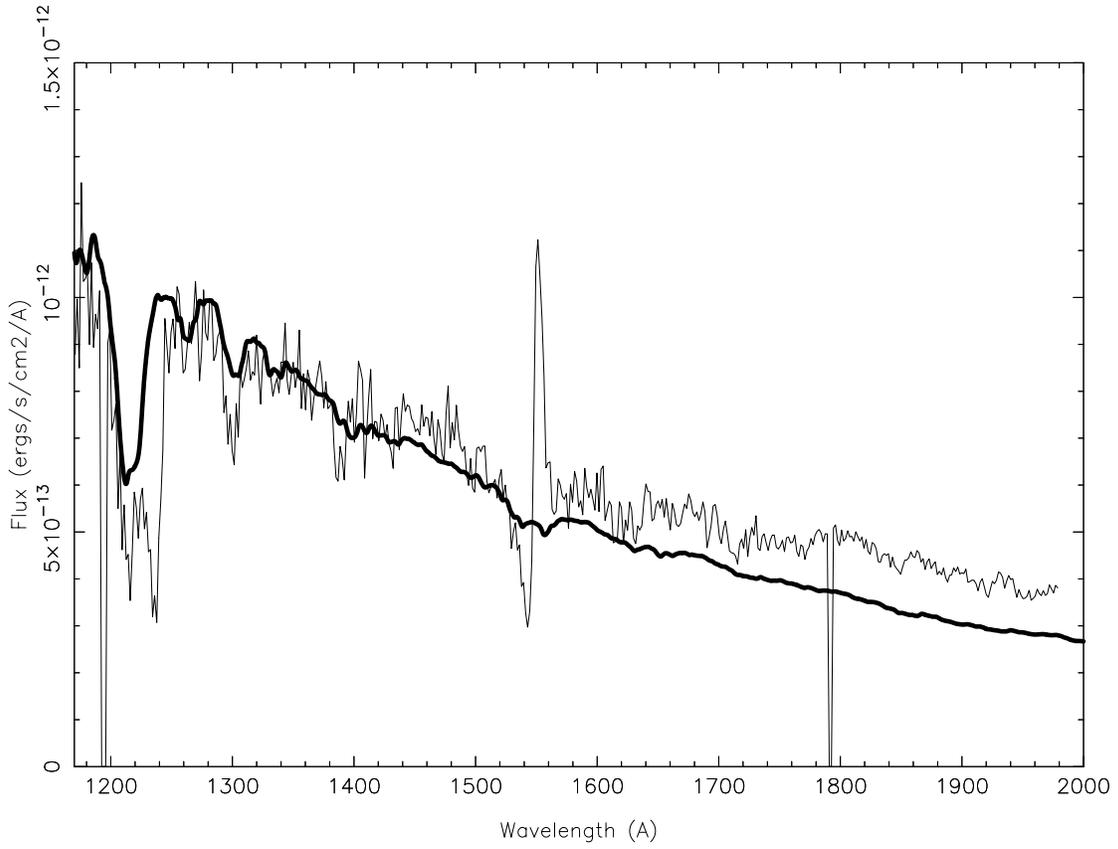}
\caption{The best fitting accretion disk to the {\it IUE} spectrum SWP15079 
for RU Peg, having a white dwarf mass of 1.21 M$_{\sun}$, an accretion rate 
of $10^{-9}$ M$_{\sun}$ yr$^{-1}$, and an inclination of 41\degr.  This fit gives a 
scale-factor distance of approximately 278 pc.\label{fig05}}
\end{figure}

\clearpage

\begin{figure}
\plotone{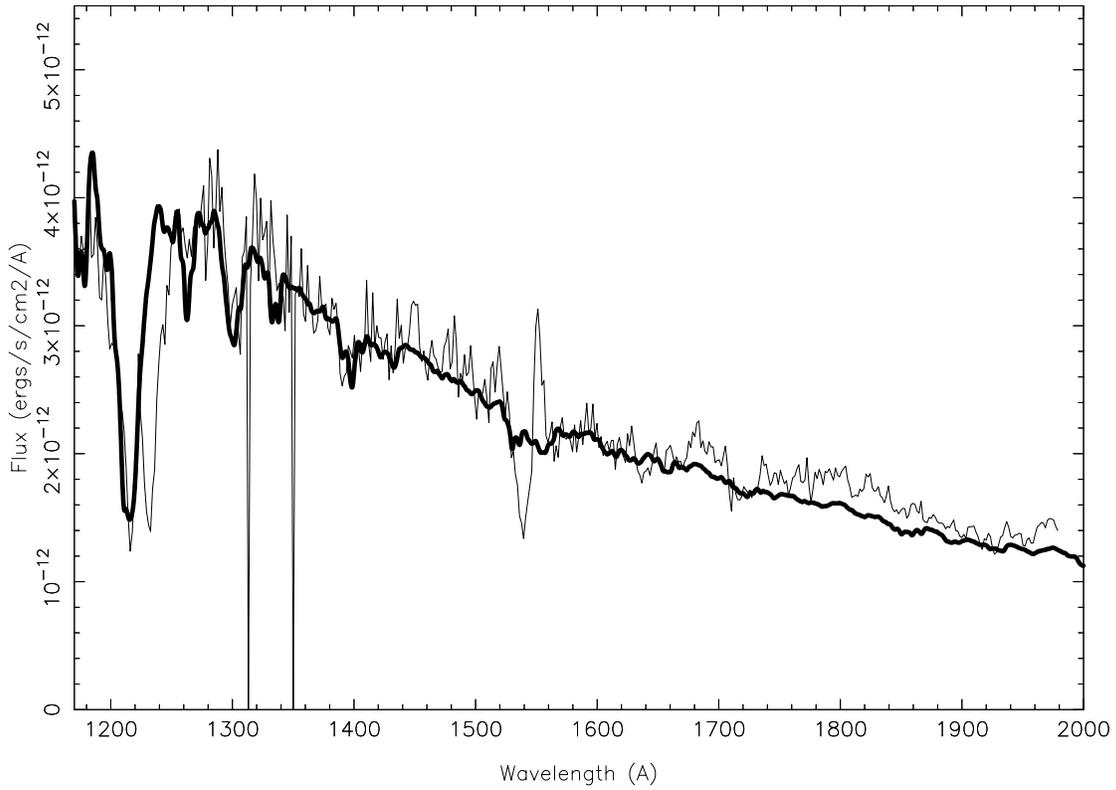}
\caption{The best fitting accretion disk to the {\it IUE} spectrum SWP33642 
for T Leo in outburst, having a white dwarf mass of 0.35 M$_{\sun}$, an 
accretion rate of $10^{-8}$ M$_{\sun}$ yr$^{-1}$, and an inclination of 60\degr.  
This fit gives a scale-factor distance of approximately 113 pc.\label{fig06}}
\end{figure}

\clearpage

\begin{figure} 
\plotone{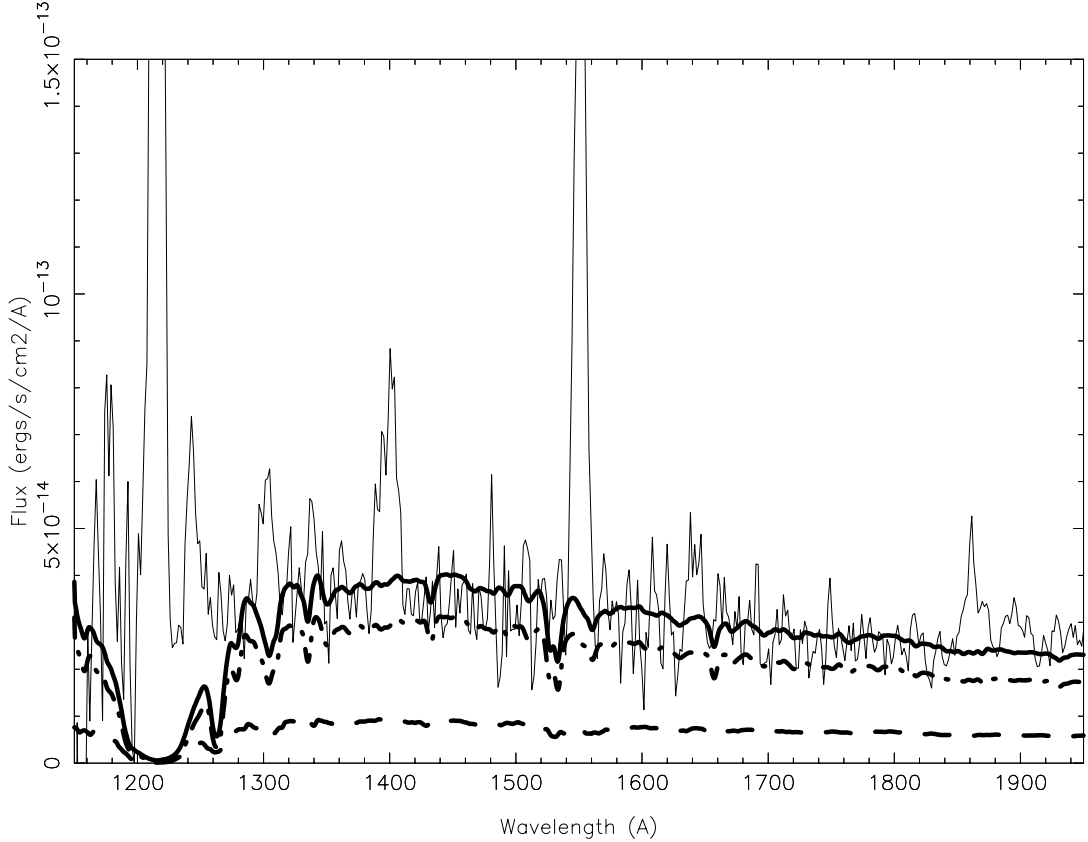} 
\caption{The best fitting combination of an accretion
disk and a white dwarf photosphere to the {\it IUE} spectrum SWP33699 for T Leo in
quiescence.  The parameters for this fit are a white dwarf mass of 0.35 M$_{\sun}$,
T$_{eff}$ of 16,000 K, $\log(g)$ = 7.55, an inclination of 60\degr, and a scaled
accretion rate of $6\times 10^{-11}$ M$_{\sun}$ yr$^{-1}$.  The white dwarf (dash-dotted
line) is seen to contribute the majority of the flux, 77\%, as would be expected during
quiescence rather than being dominated by the disk (dashed line).  This combination fit
gives a scale-factor distance of 101 pc.\label{fig07}} 
\end{figure}

\clearpage

\begin{deluxetable}{lccc}
\tablewidth{0pc}
\tablecaption{System Parameters}
\tablehead{
\colhead{System Name}&\colhead{VY Aqr}&\colhead{RU Peg}&\colhead{T Leo}
}
\startdata
Subtype& SU UMa&U Gem&SU UMa\\
P$_{orb}$ (d)& 0.06309 \tablenotemark{a} &0.3746 \tablenotemark{f} &0.05882 \tablenotemark{i}\\
$\pi$ (MAS)&$10.2\pm 1.4 $\tablenotemark{b} & $3.55\pm 0.26$ \tablenotemark{g}&$9.1\pm 0.7$ \tablenotemark{b}\\
Distance (pc)&$97\pm 13$& $282\pm 20$&$101\pm 12$\\
$i$ (\degr)&30 - 40 \tablenotemark{c} &41 \tablenotemark{g} &46 - 84 \tablenotemark{d}\\
$M_{wd}$ (M$_{\sun}$)& 0.6 - 1.2 \tablenotemark{c}&1.21 \tablenotemark{h}& 0.35 - 0.4 \tablenotemark{i}\\
$\tau_{rec}$ (d) \tablenotemark{d}&350&76&420\\
V$_{max}$ \tablenotemark{e}&8.4 &9.0 &10.0B\\
V$_{min}$ \tablenotemark{e}&17.2&13.2&15.7B\\
\enddata
\tablerefs{
           (a) \citealt{ThorstensenTaylor1997}; (b) \citealt{Thorstensen2003}; (c) \citealt{Aug1994}; (d) \citealt{Ritter2003}; (e) \citealt{DS1993}; (f) \citealt{Stover1981}; (g) \citealt{Harrison2004}; (h) \citealt{Shafter1984}; (i) \citealt{ShafterSzkody1984}
          }
\end{deluxetable}

\begin{deluxetable}{lccc}
\tablewidth{0pc}
\tablecaption{ {\it IUE} Observing Log}
\tablehead{
\colhead{Name}&\colhead{ {\it IUE} Spectrum}&\colhead{Date of Observation}&\colhead{
Exposure Length (s)}}
\startdata
VY Aqr  &       SWP21720 (Sm. Aperture)          &       12/8/1983       &       1200\\
RU Peg  &       SWP15062                         &       9/20/1981       &       600\\
        &       SWP15079                         &       9/22/1981       &       840\\
T Leo   &       SWP33642 (Outburst)              &       5/25/1988       &       300\\
        &       SWP33699 (Quiescence)            &        6/3/1988       &       2100\\
\enddata
\end{deluxetable}

\begin{deluxetable}{cccr}
\tablewidth{0pc}

\tablecaption{WZ Sge-like Temperature Comparison}
\tablehead{
\colhead{System}&\colhead{P (hours)}&\colhead{T$_{wd}$ (K)}&\colhead{Distance (pc)}
}
\startdata
WZ Sge & 1.30 &           14,800&           43\\
AL Com& 1.36    &       16,300&         845\\
SW Uma& 1.36    &       14,000&         159\\
T Leo   &1.41&          16,000&         101\\
BC Uma& 1.52&           15,200&         287\\
\enddata
\tablerefs{Adapted from \citealt{Szkody2002}}
\end{deluxetable}

\begin{deluxetable}{lccccc}
\tablewidth{0pc}
\tablecaption{Summary of Best-Fitting Models}
\tablehead{
\colhead{System}&\colhead{M$_{wd}$}&\colhead{$\dot{M}$}&\colhead{\it i}&\colhead{Distance}\\
&\colhead{(M$_{\sun}$)}&\colhead{(M$_{\sun}$ yr$^{-1}$)}&\colhead{(\degr)}&\colhead{(pc)}
}
\startdata
VY Aqr (Outburst):& 0.55&   $10^{-9}$&              41&       93&\\
RU Peg (Outburst):& 1.21&   $10^{-9}$&              41&       278&\\
T Leo (Outburst): & 0.35    &$10^{-8}$&             60&       113&\\
                  &         &                  &WD T$_{eff}$ (K)&Flux Contrib.&Dist. (pc)\\\cline{4-6}
T Leo (Quiescence):& 0.35    &$6 \times 10^{-11}$&   16,000  &       77\% WD      & 101\\
                   &         &                   &           &       23\% Disk&\\
\enddata
\end{deluxetable}       

\end{document}